\documentclass{moriond}
\usepackage{amsmath}

\def\be{\begin{equation}}
\def\ee{\end{equation}}
\def\bea{\begin{eqnarray}}
\def\eea{\end{eqnarray}}

\newcommand{\Photo}{}

\begin{document}
\vspace*{4cm}
\title{Recent Highlights from STAR BES Phase II}

\author{Dylan Neff on behalf of the STAR Collaboration}

\address{
IRFU, CEA, Université Paris-Saclay\\
91191 Gif-sur-Yvette, France}

\maketitle\abstracts{
The second phase of the RHIC Beam Energy Scan (BES-II) was conducted between 2019 and 2021. High statistics data was collected by the STAR experiment for Au+Au collisions at $\sqrt{s_{NN}}$ from 7.7 to 27 GeV in collider mode and from 3 to 13.7 GeV in fixed target mode. A selection of results from the various BES-II analyses are presented here to showcase the wide range of physics accessible.
}

\section{Introduction}

The second phase of the Beam Energy Scan at RHIC (BES-II) was performed to provide high statistics data for Au+Au collisions at low center-of-mass energies. These low energy collisions produce systems of high baryon chemical potential, $\mu_B$, up to 420 MeV in collider mode and 750 MeV in fixed target mode. The Beam Energy Scan extends experimental reach further into under-explored areas of the QCD phase diagram where the nature of the QCD phase transition can be studied.


The STAR experiment is well positioned to take full advantage of RHIC's energy scan. With full azimuthal coverage over a large pseudo-rapidity along with excellent particle identification capabilities, a wide range of physics can be probed. For BES-II, STAR has added the inner Time Projection Chamber (iTPC) to extend the mid-rapidity acceptance along with the endcap Time Of Flight (eTOF) to enhance particle identification in fixed target mode. The Event Plane Detector (EPD) was also added for event plane estimation using forward-going particles. We discuss here a subset of the ongoing STAR analyses on BES-II data.

\begin{figure}[bh]
\centerline{\includegraphics[width=\linewidth]{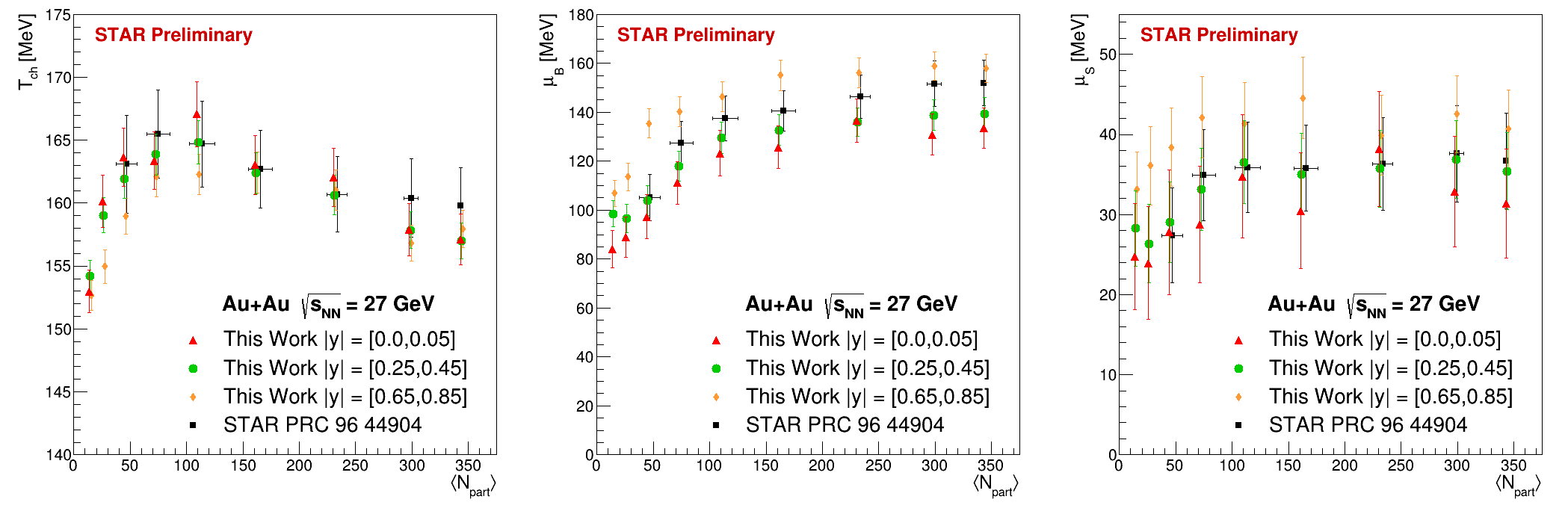}}
\caption[]{Thermodynamic properties of the medium produced in Au+Au collisions at $\sqrt{s_{NN}}$ = 27 GeV are extracted from measured yields and plotted as a function of the number of participants. STAR BES-I data at mid-rapidity are shown in black, while three rapidity ranges are plotted for BES-II data.}
\label{fig:chem_freezeout}
\end{figure}

\section{Rapidity Dependence of Chemical Freeze-Out}
\label{subsec:chem_freezeout}

As RHIC scans $\sqrt{s_{NN}}$, it is imperative to understand where on the QCD phase diagram the produced QGP systems lie. Thermodynamic properties of the system at chemical freeze-out can be extracted by measuring particle yields and comparing to the expectations from a thermal model (Thermus~\cite{thermus}). Extracted temperature along with baryon and strange chemical potential are shown in Figure~\ref{fig:chem_freezeout} for BES-I data at 27~GeV~\cite{star_bes1_yeilds}.

The extracted parameters from BES-II 27~GeV data are shown in Figure~\ref{fig:chem_freezeout} for three rapidity ranges. The variation between these regions, $\Delta \mu_B \approx \text{25 MeV}$ and $\Delta \mu_S \approx \text{10 MeV}$ for $\Delta y = 1$, helps to understand how the properties of the system change with rapidity.

\section{Thermal Dielectrons}
\label{subsec:thermal_dielectrons}

\begin{figure}[b]
\begin{minipage}{0.47\linewidth}
\centerline{\includegraphics[width=\linewidth]{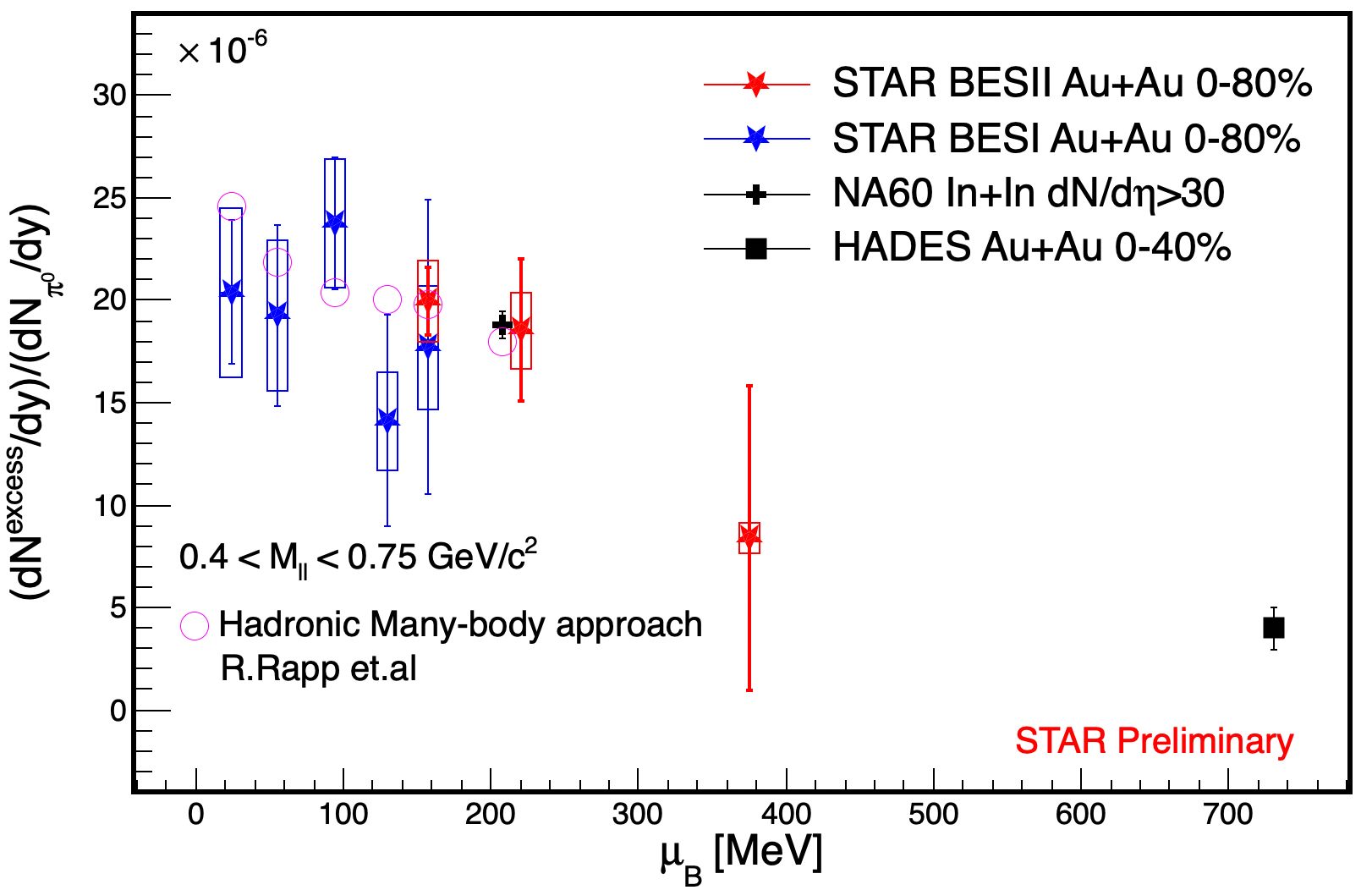}}
\end{minipage}
\hfill
\begin{minipage}{0.47\linewidth}
\centerline{\includegraphics[width=\linewidth]{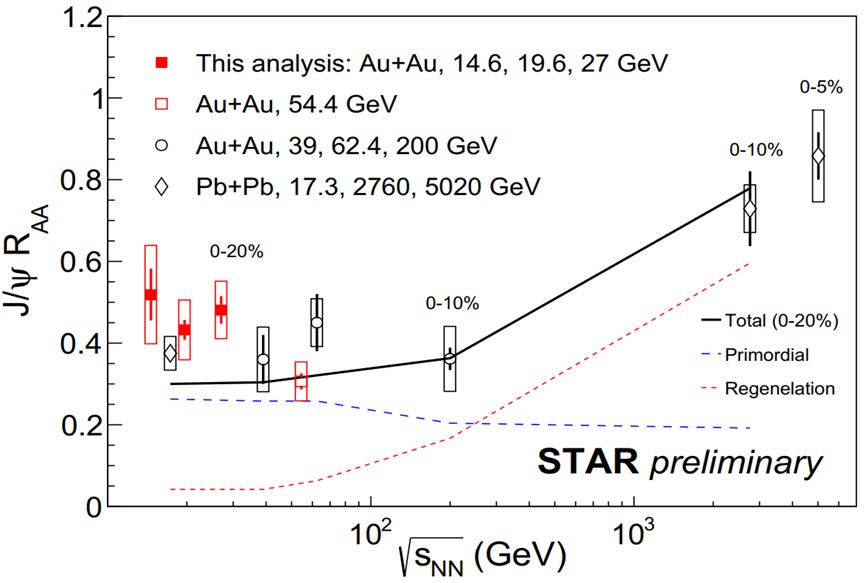}}
\end{minipage}
\caption[]{(left) The excess yield of dielectrons is plotted vs baryon chemical potential, $\mu_B$. STAR BES-II data is shown in red for 7.7, 14.6 and 19.6~GeV. (right) J$/\Psi$ $\text{R}_{\text{AA}}$ is plotted as a function of center-of-mass collision energy. STAR BES-II data for 14.6, 19.6 and 27~GeV is plotted along with theoretical curves depicting the contribution from direct suppression (primordial) and regeneration\cite{jpsi_model}.}
\label{fig:thermal_dielectrons_jspi}
\end{figure}

While measured particle yields can provide information about the thermodynamic properties of the QGP system at chemical freeze-out, thermal dielectrons are able to probe these properties earlier in the QGP evolution. Electron positron pairs are identified and their yield is corrected for known physical processes. What remains (excess yield) are thermal dielectrons, originating from thermally radiated photons. Photons are not influenced by the strong forces at play in the QGP evolution and are therefore ideal probes of the evolving fireball. 

The excess dielectron yield is measured in STAR BES-II data at 19.6, 14.6 and 7.7~GeV (Figure~\ref{fig:thermal_dielectrons_jspi} left). While error bars on the 7.7~GeV point are large, the trend hints at a decreasing yield with increasing $\mu_B$.

\section{Energy Dependence of J$/\Psi$ $\text{R}_{\text{AA}}$}
\label{subsec:jpsi}

The normalized yield of J$/\Psi$ in Au+Au collisions compared to that of p+p collisions ($\text{R}_{\text{AA}}$) is another observable whose energy dependence is of interest. J$/\Psi$ mesons produced in heavy ion collisions undergo interactions with the hot QGP medium as it evolves. The dominant effect of this interaction is the melting of the meson into free $c$ and $\bar{c}$ quarks, which hadronize with other quarks from the medium and suppress the yield in Au+Au when compared to p+p. However, the number of charm quarks produced in Au+Au collisions and therefore the charm density increases with collision energy, increasing the probability of recombining dissociated pairs. Together, these effects are expected to produce a  J$/\Psi$ $\text{R}_{\text{AA}}$ which increases with energy.

STAR has measured the J$/\Psi$ $\text{R}_{\text{AA}}$ in BES-II data at 14.6, 19.6 and 27~GeV. Combined with BES-I results~\cite{star_bes1_jpsi}, no significant energy dependence is observed in central collisions from 14.6 to 200~GeV, within uncertainties.

\section{Disappearance of NCQ Scaling}
\label{subsec:ncq_scaling}

\begin{figure}
\begin{minipage}{0.46\linewidth}
\centerline{\includegraphics[width=\linewidth]{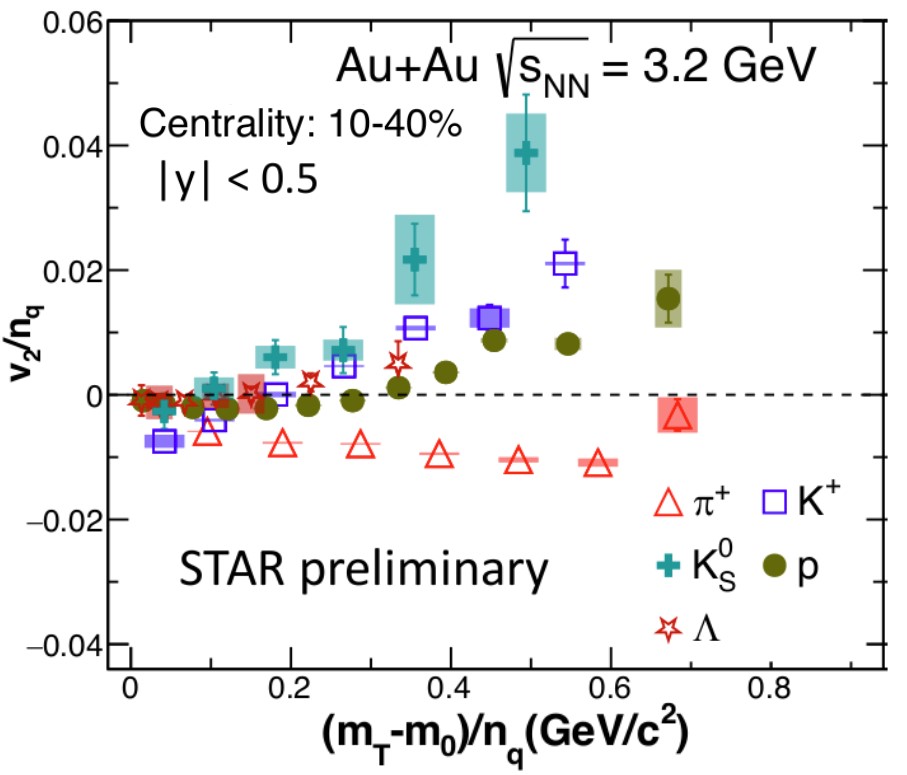}}
\end{minipage}
\hfill
\begin{minipage}{0.42\linewidth}
\centerline{\includegraphics[width=\linewidth]{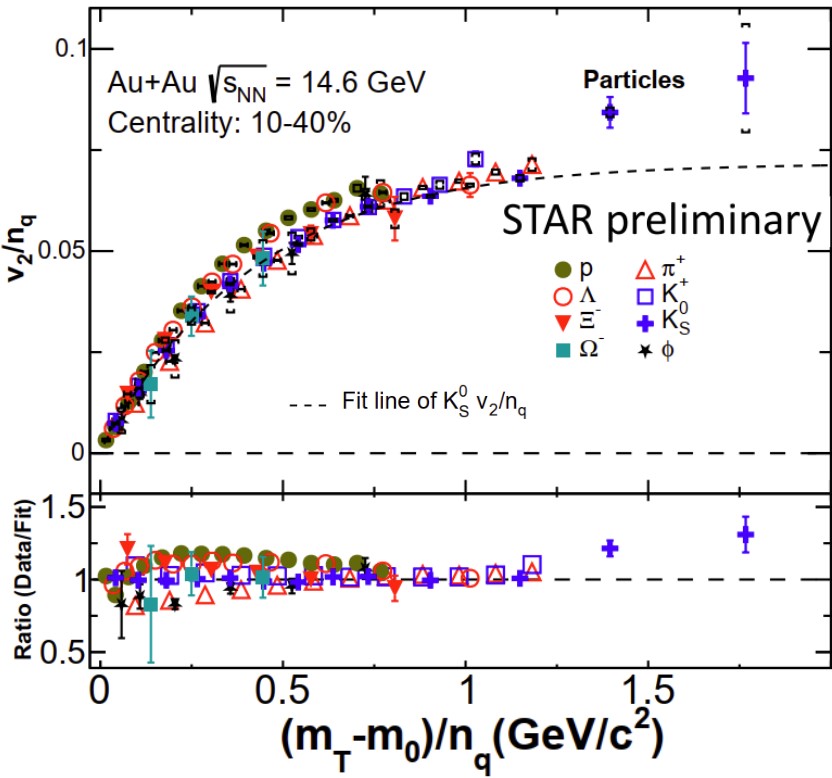}}
\end{minipage}
\caption[]{$v_2/n_q$ of identified particles is plotted vs the transverse mass per constituent quark for Au+Au collisions at $\sqrt{s_{NN}}$ = 3.2 GeV on the left and 14.6~GeV on the right.}
\label{fig:ncq_scaling}
\end{figure}

Plotting the elliptic flow ($v_2$) of identified hadrons divided by their number of constituent quarks (two for mesons, three for baryons) against their transverse mass per quark, a universal trend is observed at 14.6 GeV in BES-II data (Figure~\ref{fig:ncq_scaling} right). This number of constituent quark (NCQ) scaling is observed also at higher collision energies at RHIC and is interpreted as the result of independent quark degrees of freedom flowing in the QGP medium and coalescing into hadrons at chemical freeze-out. NCQ scaling is observed to disappear in STAR's BES-II fixed target 3.2~GeV data (Figure~\ref{fig:ncq_scaling} left). This could indicate a lack of partonic degrees of freedom, suggesting the dominance of hadronic interactions at this energy. 

\section{Search for the Chiral Magnetic Effect}\label{subsec:cme}

A net charge separation along the direction of the magnetic field is expected in systems with an imbalance of charged chiral particles and a strong magnetic field, a phenomena known as the Chiral Magnetic Effect (CME). 
Heavy ion collisions are expected to produce extreme magnetic fields as the charged spectators move away from the fireball, and gluon excitation can produce a chirality imbalance in local topological domains. 
STAR searches for CME in BES-II data via the $\Delta \gamma^{112}$ observable, which is the difference between $\gamma^{112}\equiv \langle \cos \left( \varphi_{\alpha} + \varphi_{\beta} - 2 \Psi_{\text{RP}} \right) \rangle$ of opposite-sign and same-sign pairs of particles ($\varphi_{\alpha,\beta}$ are the azimuthal angles of paired particles and $\Psi_{\text{RP}}$ is the reaction plane angle). A novel Event Shape Selection \cite{XU2024138367} (ESS) method is employed to mitigate the elliptic flow background. The measurement also uses the event plane from EPD spectator plane to suppress the nonflow background.

Figure~\ref{fig:cme_femto} (left) shows the $\Delta \gamma^{112}_{\rm ESS}$ from the intercept measured by ESS of $\Delta \gamma^{112}$ at zero elliptic flow.
A positive charge separation above $3\sigma$ is observed at 14.6 and 19.6~GeV, indicating the possibility of a CME signature at these energies.
Meanwhile, the corresponding background indicator $\Delta \gamma^{132}_{\rm ESS}$ is found to be consistent with zero.

\section{Femtoscopy}\label{subsec:femto}

\begin{figure}
\begin{minipage}{0.50\linewidth}
\centerline{\includegraphics[width=\linewidth]{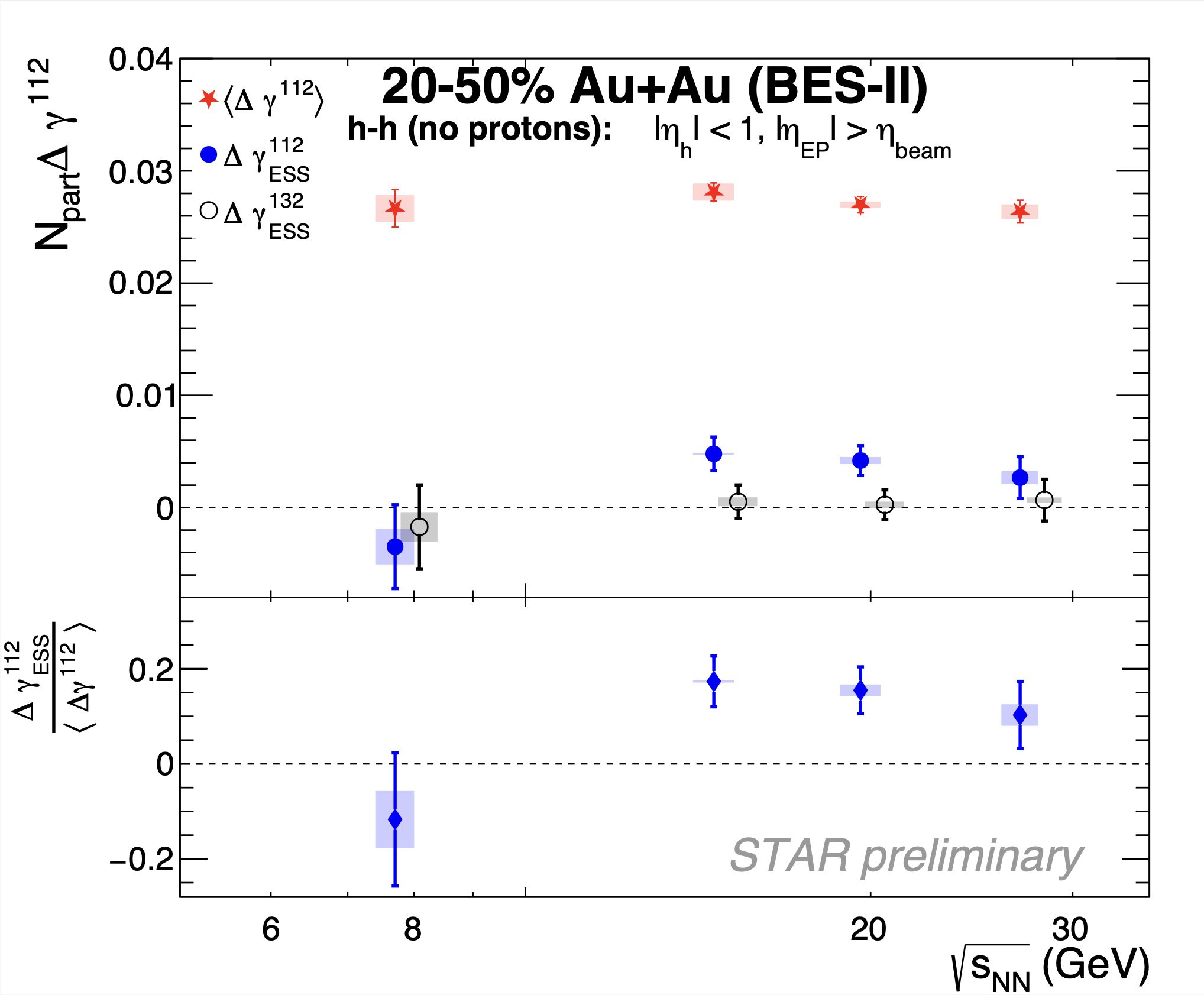}}
\end{minipage}
\hfill
\begin{minipage}{0.50\linewidth}
\centerline{\includegraphics[width=\linewidth]{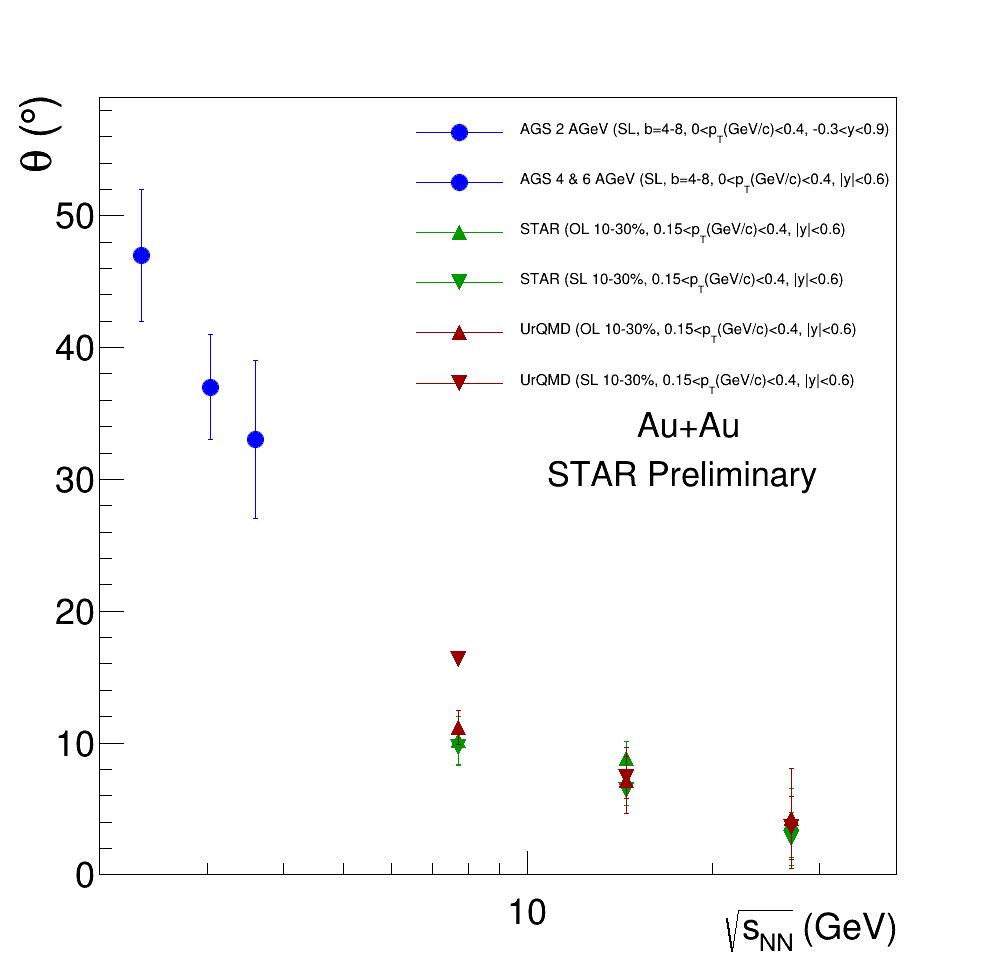}}
\end{minipage}
\caption[]{(left) $\text{N}_{\text{part}} \Delta \gamma^{112}$ is plotted as a function of collision energy before (red) and after (blue) event shape selection (ESS). A significantly positive signal is observed at 14.6 and 19.6~GeV. (right) The tilt of the fireball is extracted and presented as a function of energy.}
\label{fig:cme_femto}
\end{figure}

The size and orientation of the produced QGP fireball is studied in BES-II data via momentum space correlations between emitted particles. Along with the three dimensional size of the fireball, its tilt with respect to the beam axis has also been measured at RHIC for the first time in BES-II data at 7.7, 14.6 and 27~GeV. In Figure~\ref{fig:cme_femto} (right) the tilt is found to increase with decreasing $\sqrt{s_{NN}}$, connecting with AGS data at 2, 4 and 6~GeV.

\section{Summary}

The second phase of the RHIC Beam Energy Scan (BES-II) was successfully conducted between 2019 and 2021, producing high statistics data for Au+Au collisions at $\sqrt{s_{NN}}$ from 3 GeV to 27 GeV. The STAR collaboration is currently analyzing this data to explore a wide range of physics topics. Early highlights include a study of the rapidity dependence of chemical freeze-out, which quantifies the sensitivity of baryon ($\mu_B$) and strange ($\mu_S$) chemical potentials to rapidity variation. Thermal dielectron yields probe the temperature of the QGP medium through its evolution, with BES-II results hinting at a decreasing trend with increasing $\mu_B$. The J$/\Psi$ $\text{R}_{\text{AA}}$ measurement reflects the interplay between charmonium dissociation and recombination in the hot QGP medium and shows no significant energy dependence from 14.6 to 200 GeV. The disappearance of NCQ scaling at 3.2 GeV suggests the absence of partonic degrees of freedom, indicating a dominance of hadronic interactions at this energy. The search for the Chiral Magnetic Effect using a novel background mitigation technique shows potential evidence for charge separation between 14.6 and 19.6 GeV. Finally, the tilt of the fireball produced in Au+Au collisions has been measured via femtoscopic correlations and is found to increase as the collision energy decreases.




\section*{References}
\bibliography{references}






\end{document}